\documentclass[journal,twoside,web]{ieeecolor}
\usepackage{tmi}
\usepackage{cite}
\usepackage{array}
\usepackage{tabularx}
\usepackage{amsmath,amssymb,amsfonts}
\usepackage{booktabs}
\newtheorem{lemma}{Lemma}[section]
\usepackage{algorithmic}
\usepackage{ragged2e}  
\usepackage{multirow}
\usepackage{amsfonts} 
\usepackage{graphicx}
\usepackage{float}
\usepackage{xcolor}
\usepackage{textcomp}
\usepackage{longtable}
\definecolor{lightblue}{RGB}{173,216,230}
\definecolor{myblue}{RGB}{0,0,255}
\usepackage{hyperref}
\hypersetup{
	hypertex=true,
	colorlinks=true,
	linkcolor=myblue,
	anchorcolor=myblue,
	citecolor=myblue
}
\usepackage{caption} 
\def\BibTeX{{\rm B\kern-.05em{\sc i\kern-.025em b}\kern-.08em
		T\kern-.1667em\lower.7ex\hbox{E}\kern-.125emX}}
\begin{document}
\title{
	Visible Singularities Guided Correlation Network for Limited-Angle CT Reconstruction
}
\author{Yiyang Wen, Liu Shi, Zekun Zhou, WenZhe Shan, Qiegen Liu, \IEEEmembership{Senior Member, IEEE}\vspace{-2.5 em}
	\thanks{This study was funded by National Natural Science Foundation of China (621220033), Early Stage Young Scientific and Technological Talent Training Foundation of Jiangxi Province (Grant: 20252BEJ730005), Nanchang University Youth Talent Training Innovation Fund Project(Grant: XX202506030012). (Y. Wen and L. Shi are co-first authors) (Co-corresponding authors: L. Shi and Q. Liu.)}
	\thanks{This work was supported by data from YOFO Technology Co. Ltd. Hefei, 230088, China(info@yofomedical.com).}
	\thanks{This work was supported by data from the Institute of Jinan Laboratory of Applied Nuclear Science.}
	\thanks{Y. Wen, L. Shi, W. Shan and Q. Liu are with the School of Information Engineering, Z. Zhou is with School of Mathematics
	and Computer Sciences , Nanchang University, Nanchang 330031,
	China (email: \{shiliu, shan, liuqiegen\}@ncu.edu.cn, \{416100240092, ZekunZhou\}@email.ncu.edu.cn).}
}
\markboth{Journal of \LaTeX\ Class Files,~Vol.~14, No.~8, August~2021}%
{Shell \MakeLowercase{\textit{et al.}}: A Sample Article Using IEEEtran.cls for IEEE Journals}
\maketitle

\begin{abstract}
Limited-angle computed tomography (LACT) offers the advantages of reduced radiation dose and shortened scanning time. Traditional reconstruction algorithms exhibit various inherent limitations in LACT. Currently, most deep learning-based LACT reconstruction methods focus on multi-domain fusion or the introduction of generic priors, failing to fully align with the core imaging characteristics of LACT—such as the directionality of artifacts and directional loss of structural information, which are caused by the absence of projection angles in certain directions. Inspired by the theory of visible and invisible singularities, taking into account the aforementioned core imaging characteristics of LACT, we propose a Visible Singularities Guided Correlation network for LACT reconstruction (VSGC). The design philosophy of VSGC consists of two core steps: First, extract VS edge features from LACT images and focus the model's attention on these VS. Second, establish correlations between the VS edge features and other regions of the image. Additionally, a multi-scale loss function with anisotropic constraint is employed to constrain the model to converge in multiple aspects. Finally, qualitative and quantitative validations are conducted on both simulated and real datasets to verify the effectiveness and feasibility of the proposed design. Particularly, in comparison with alternative methods, VSGC delivers more prominent performance in small angular ranges, with the PSNR improvement of 2.45 dB and the SSIM enhancement of 1.5\%. The code is publicly available at \url{https://github.com/yqx7150/VSGC}.

\end{abstract}


\section{Introduction}
\IEEEPARstart{i}{n} X-ray CT imaging for medical and other fields, full-angle projection data cannot always be acquired due to practical constraints such as equipment performance and properties of the imaged object (e.g., breast tomosynthesis\cite{niklason1997digital}, C-arm neuroimaging\cite{fahrig2006dose}, image-guided surgery\cite{bachar2007image}). Limited-angle CT (LACT) reconstruction technology enables tomographic imaging based on limited-angle projection data\cite{hess2014pivotal}. The reconstruction of LACT using algorithms such as FBP\cite{ramachandran1971three} and FDK\cite{feldkamp1984practical} exhibits severe artifacts\cite{de2018gpu,quinto1993singularities,quinto2017artifacts}. Although traditional iterative reconstruction algorithms\cite{kak2001principles,andersen1984simultaneous,trampert1990simultaneous} exhibit a certain degree of adaptability to incomplete data, they generally suffer from inherent bottlenecks, including slow convergence, noise sensitivity, high computational complexity, and over-smoothed image structures\cite{zhu2014iterative}.


\begin{figure}[!t]
	\centering
	\includegraphics[width=0.9\columnwidth]{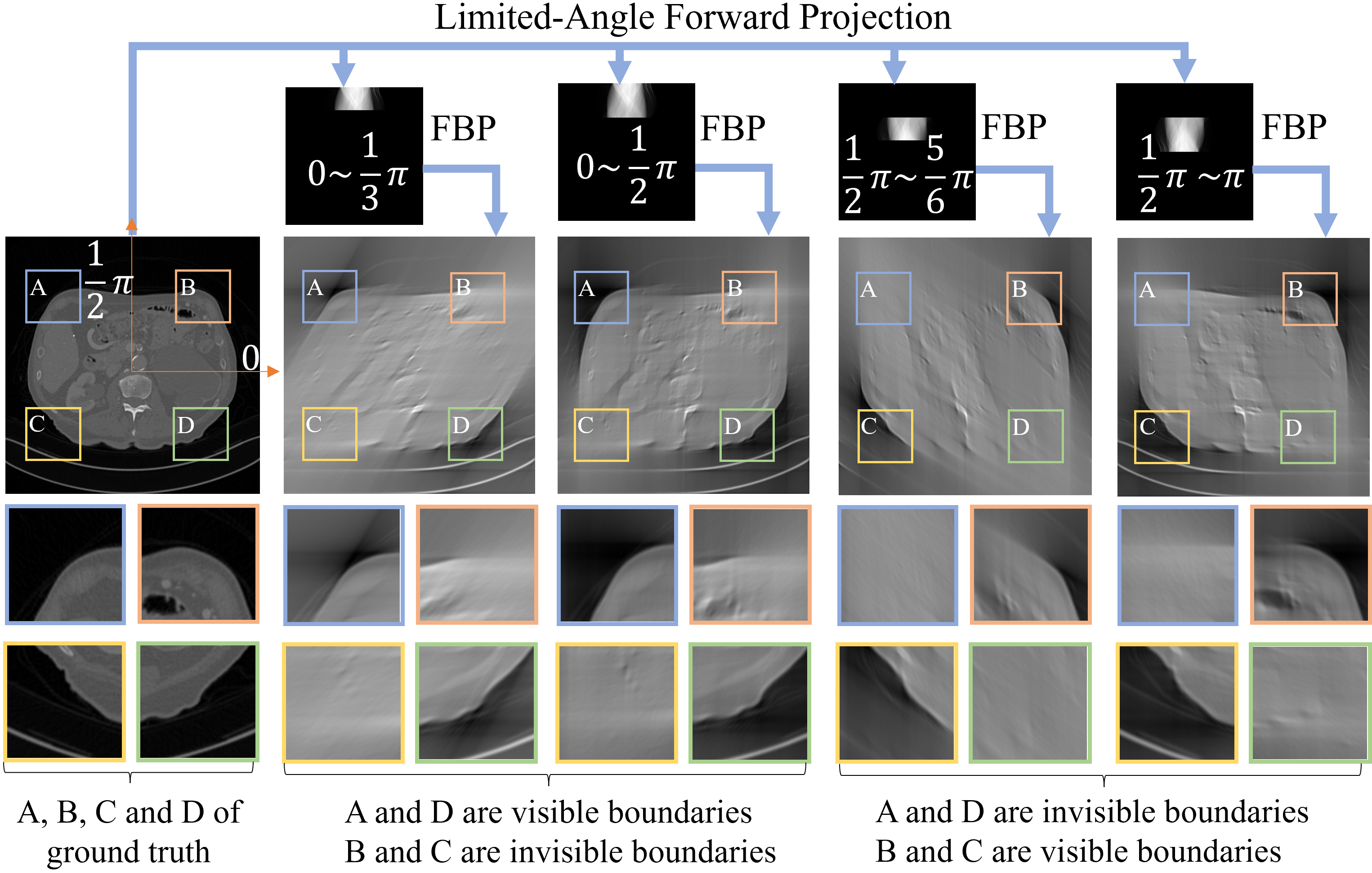} 
	\caption{LACT boundary reconstructibility correlate with parallelism between boundary tangent directions and X-ray propagation directions. If the tangent line of a body feature's boundary is parallel to a particular scan ray, that boundary will be easily reconstructed. If the tangent line of a body feature's boundary is not parallel to any scan ray, that boundary will be difficult to reconstruct\cite{quinto1993singularities}.}
	\label{limited-angle}
	\vspace{-15pt}
\end{figure}

Substantial efforts in regularization design have been made rencently for iterative algorithms for LACT: Wang \textit{et al}. \cite{wang2017reweighted} proposed the method of Anisotropic Total Variation (ATV) which incorporates the angular range of projection data as additional prior information by assigning direction-dependent weights. Zhang \textit{et al}. \cite{zhang2021directional}formulated the LACT reconstruction problem as a convex optimization problem imposing Directional Total Variation (DTV) constraints on images. Xu \textit{et al}. \cite{xu2019image} proposed AEDS algorithm which imposes constraints on the partial derivatives of images along two orthogonal directions. All the aforementioned designs adopt direction-specific constraints which require performing directional optimization on traditional TV\cite{rudin1992nonlinear} regularization to adapt to the characteristics of limited-angle data (LAD). However, complex constraints can lead to increased computational cost, elevated difficulty in parameter tuning, and insufficient generalization ability\cite{zhang2025pp}.


Deep learning-driven LACT reconstruction methods are mainly implemented through three types of strategies: 1) integration with iterative optimization (e.g., DIOR\cite{hu2022dior}, IRON\cite{pan2024iterative}, MIST-Net\cite{pan2022multi}), 2) fusion with traditional reconstruction algorithms (e.g., MSDDRNet\cite{zhou2023multi}), and 3) reliance on generative models (e.g., SIAR-GAN\cite{xie2022limited}, WISM\cite{zhang2024wavelet}, DOLCE\cite{liu2023dolce}). By exploring the statistical features of multi-domain data (including image, projection, and wavelet domains) and introducing constraints such as data consistency and physical priors, these methods can effectively suppress artifacts caused by LAD and restore fine image details. However, most existing approaches focus on multi-domain fusion or the introduction of universal priors, and fail to fully conform to the core imaging characteristics of LACT, such as the directionality of artifacts and directional loss of structural information induced by the absence of projection angles in certain directions. As shown in Fig. \ref{limited-angle}, the reconstructibility performance of LACT boundaries is dependent on the parallelism between boundary tangent directions and X-ray propagation directions.


Quinto \textit{et al}. \cite{quinto1993singularities} proposed the theory of Visible Singularities (VS) and Invisible Singularities (IVS), where the VS is defined as the parts of the object boundary tangent to the rays in the LAD, while the IVS refers to the parts of the object boundary with no rays from the LAD tangent to them. They proved that the VS can be accurately reconstructed, capturing significant signal mutations through measured data. In contrast, the signal characteristics of the IVS are smoothed and weakened in the LAD, making direct recovery difficult. Inspired by the VS/IVS theory, we propose a deep learning reconstruction network tailored for the directional artifacts and directional loss of structural information in LACT: the Visible Singularities Guided Correlation network (VSGC). Our design consists of two core steps: 1) Accurate extraction of the edge features of VS from LACT images, and 2) establishing correlations between these VS edge features and other regions. In addition, a multi-scale loss function with anisotropic constraint has been designed to constrain the model convergence from multiple aspects.
   

Our main contributions are as follows:
\begin{itemize}
	\item \textbf{Targeted VS Edge Feature Capture.} To enhance the model's ability to extract VS edge features, we designed the visible singularities wavelet dense (VSWD) which performs edge fusion to enhance the VS edge features of the model input and utilizes three layers of encoding units embedded with wavelet convolution to focus on extraction of the VS edge features from the intermediate feature maps of the model.
	
	\item \textbf{VS Edge Feature Inter-Regions Linkage.} To establish correlations between the edge features of VS and other image regions, we proposed a multi-scale visible singularity-crossregion correlation self-attention mechanism (MCA), with the attention score matrix retaining only the top-k values in each row to preserve critical global correlations, and convolutional channels to capture detailed local correlations within image patches.
	
	\item \textbf{Multi-Scale Loss Function with Anisotropic Constraint.} We developed a novel multi-scale loss function with four integrated components: an anisotropic weighted part to prioritize regions severely affected by LACT artifacts, a structural similarity (SSIM) part to enhance the visual perception of reconstructed images, an edge gradient part to improve the clarity of image textures and edges, and a perceptual part to alleviate the over-smoothing issue of reconstructed images.
\end{itemize}


The structure of this paper is organized as follows: Section \ref{sec:related_work} is devoted to the reconstruction of LACT as well as its artifacts and singularity characteristics. Section \ref{Method} presents a comprehensive elaboration of the proposed VSGC model. Section \ref{experiments} demonstrates the reconstruction performance of different methods through comparative analysis. Section \ref{discussion} discusses relevant issues. Section \ref{conclusion} summarizes the conclusions.

\section{Related Work}
\label{sec:related_work} 


\subsection{LACT Reconstruction}

CT imaging consists of two steps: data acquisition and reconstruction. For the data acquisition, X-ray projections of the object are captured from different viewing angles. The attenuation coefficient distribution function \( f: \mathbb{R}^2 \to \mathbb{R} \) of the material is then reconstructed from the projection data. We adopt the Radon transform as the mathematical model for the data acquisition step, which is defined as follows:
\begin{equation}
	\mathcal{R}f(\theta, s) = \iint_{L(\theta,s)} f(x,y) \, dxdy,
\end{equation}
where \( \theta \in [0, \pi) \) represents the angle between the incident direction of the X-ray beam and the positive \( x \)-axis. The equation of the line \( L(\theta, s) \) is \( x\cos\theta + y\sin\theta = s \). The FBP is a classic algorithm for CT reconstruction, where the projection data \( \mathcal{R}f \) is firstly filtered in the Fourier domain to obtain the filtered projection \( \hat{p}(\theta, s) \):
\begin{equation}
	\hat{p}(\theta, s) = \mathcal{F}^{-1}\left\{ |\omega| \cdot \mathcal{F}\{\mathcal{R}f\}(\omega) \right\}(s),
\end{equation}
where \( \mathcal{F} \) and \( \mathcal{F}^{-1} \) denote the one-dimensional Fourier transform and inverse Fourier transform (with respect to \( s \)), respectively. \( |\omega| \) is the Ram-Lak filter kernel. The filtered data is then projected back onto the 2D plane:
\begin{equation}
	f(x,y) = \frac{1}{2\pi} \int_0^\pi \hat{p}(\theta, s) \, d\theta.
\end{equation}
The FBP reconstruction can be summarized as:
\begin{equation}
	f = \mathcal{R}^* \Lambda \mathcal{R}f,
	\label{equation4}
\end{equation}
where \( \mathcal{R}^* \) is the back-projection operator, and \( \Lambda \) is the filtering operator. The limited-angle FBP reconstruction refers to directly restricting the integral range of full-angle FBP to \( [\theta_1, \theta_2] \):
\begin{equation}
	f_{l} = \frac{1}{2\pi} \int_{\theta_1}^{\theta_2} \hat{p}(\theta, s) \, d\theta,
\end{equation}
where $\theta_2-\theta_1<\pi$. Due to incomplete angular coverage, the $f_l$ reconstructed directly using the vanilla FBP suffers from severe artifacts and fails to accurately recover the distribution of attenuation coefficients.  

\subsection{Artifacts and Visible Singularities in Limited Data}
The core scientific problem of LACT resides in the reconstructibility discrepancies of object singularities and artifacts induced by data incompleteness. Relevant research has gradually established a systematic framework of characterization, using the microlocal analysis\cite{krishnan2015microlocal} and the Radon transform theory as main tools. The microlocal analysis, as a state-of-the-art mathematical branch developed on the basis of Fourier transform, has the key advantage of precisely characterizing the position, direction, and intensity properties of function singularities within the unified framework of local spatial regions and the frequency domain (cotangent space). For the characterization of singularities in microlocal analysis, the wavefront set\cite{hormander1971fourier}\cite{brouder2014smooth} served as the critical theoretical foundation for analyzing singularity propagation and distortion, as well as for the artifact generation mechanisms caused by data incompleteness.

\newtheorem{definition}{Definition} 

\begin{definition} \cite{hormander2007analysis}	
	Let \( \bar{f} \in L^2_{\text{loc}}(\mathbb{R}^2) \), i.e., \( \bar{f} \) is square integrable on every compact subset of \( \mathbb{R}^2 \). The wavefront set of \( \bar{f} \) is defined by:
	\begin{align}
		\mathrm{WF}(\bar{f}) &:= \left\{ (x_i, \xi_i) \in \mathbb{R}^2 \times \mathbb{R}^2 : \right. \nonumber \\
		&\quad \left. \bar{f} \text{ is not smooth at } x_i \text{ in the direction } \xi_i \right\},
		\label{eq:wavefront_set}
	\end{align}
	where \( x_i \in \mathbb{R}^2 \), \( i \in \mathbb{N} \) and \( \xi_i \in \mathbb{R}^2 \setminus \{ \mathbf{0} \} \). The wavefront set is a subset of the cotangent space \( T^*(\mathbb{R}^2) \).
\end{definition}
 
Quinto \cite{quinto1993singularities} was the first to establish the correlation between the X-ray transform and object singularities based on microlocal analysis, and proposed the singularity visibility criterion: If the normal singularity of the object boundary coincides with the normal direction of rays in the dataset (i.e., the boundary is tangent to the rays), such singularity is defined as VS and can be stably reconstructed. Otherwise, it is IVS, the reconstruction of which is ill-posed. The formal definitions of VS and IVS are given as follows\cite{quinto2017artifacts}: given a point-direction pair $(x_i, \xi_i)$ with $x_i \in \mathbb{R}^2$ and $\xi_i \in \mathbb{R}^2 \setminus \{ \mathbf{0} \}$, if $\xi_i$ is parallel to the normal line $\bar{\theta}(\varphi)$ of some projection angle $\varphi$, then $(x_i, \xi_i)$ is referred to as a VS of $f$, otherwise it is referred as an IVS of $f$. Quinto further proposed the geometric principle of the artifact generation: for LACT (with $\varphi \in [-\alpha, \alpha]$), streaking artifacts tend to appear along the tangent lines to the feature boundaries of the object, corresponding to the endpoints of the angular range ($\varphi = \pm \alpha$). Quinto's theory, for the first time, linked the artifact formation to data geometry and object geometry, laying a theoretical foundation for the qualitative analysis of artifacts.


\section{Method}
\label{Method}


\subsection{Motivation}

Defining the projection angle $\varphi \in [-\alpha, \alpha]$ with $\alpha \in (0, \pi/2)$, we propose a modified version of Eq. (\ref{equation4}) as:
\begin{equation}
	\mathcal{L}_\alpha f = R_\alpha^* \Lambda R f,
	\label{equation8}
\end{equation}
where $\mathcal{L}_\alpha$ denotes the limited-angle FBP operator, and $R_\alpha^*$ is the back-projection operator corresponding to the limited angle $\alpha$. For the LACT image, $\mathcal{L}_\alpha f$ exhibits the characteristics of artifact directionality and directional loss of structural information. Anisotropic analysis of $\mathcal{L}_\alpha f$ indicates that $\mathcal{L}_\alpha f$ preserves abundant structural information at the VS, where the singularities can be retained during the reconstruction process. On the other hand, $\mathcal{L}_\alpha f$ becomes blurred at the IVS, with the reconstruction process losing structural information and singularities. Moreover, streaking artifacts appear in $\mathcal{L}_\alpha f$ at $\varphi = \pm\alpha$. The anisotropy of $\mathcal{L}_\alpha f$ can be characterized via the following lemma:
    
\begin{lemma}\cite{quinto2017artifacts}
	Microlocal Regularity Theorem for $\mathcal{L}_\alpha$.
	\begin{enumerate}
		\item If $(x_i, \xi_i)$ is a VS of $f$, then $(x_i, \xi_i) \in \mathrm{WF}(\mathcal{L}_\alpha f)$.
		\item If $(x_i, \xi_i)$ is an IVS of $f$, then $(x_i, \xi_i) \notin \mathrm{WF}(\mathcal{L}_\alpha f)$.
	\end{enumerate}
	where $f$ is a function with compact support, and $\mathrm{WF}(\cdot)$ denotes the wave front set.
\end{lemma}

\begin{figure}[!t]
	\centering
	\includegraphics[width=0.9\columnwidth]{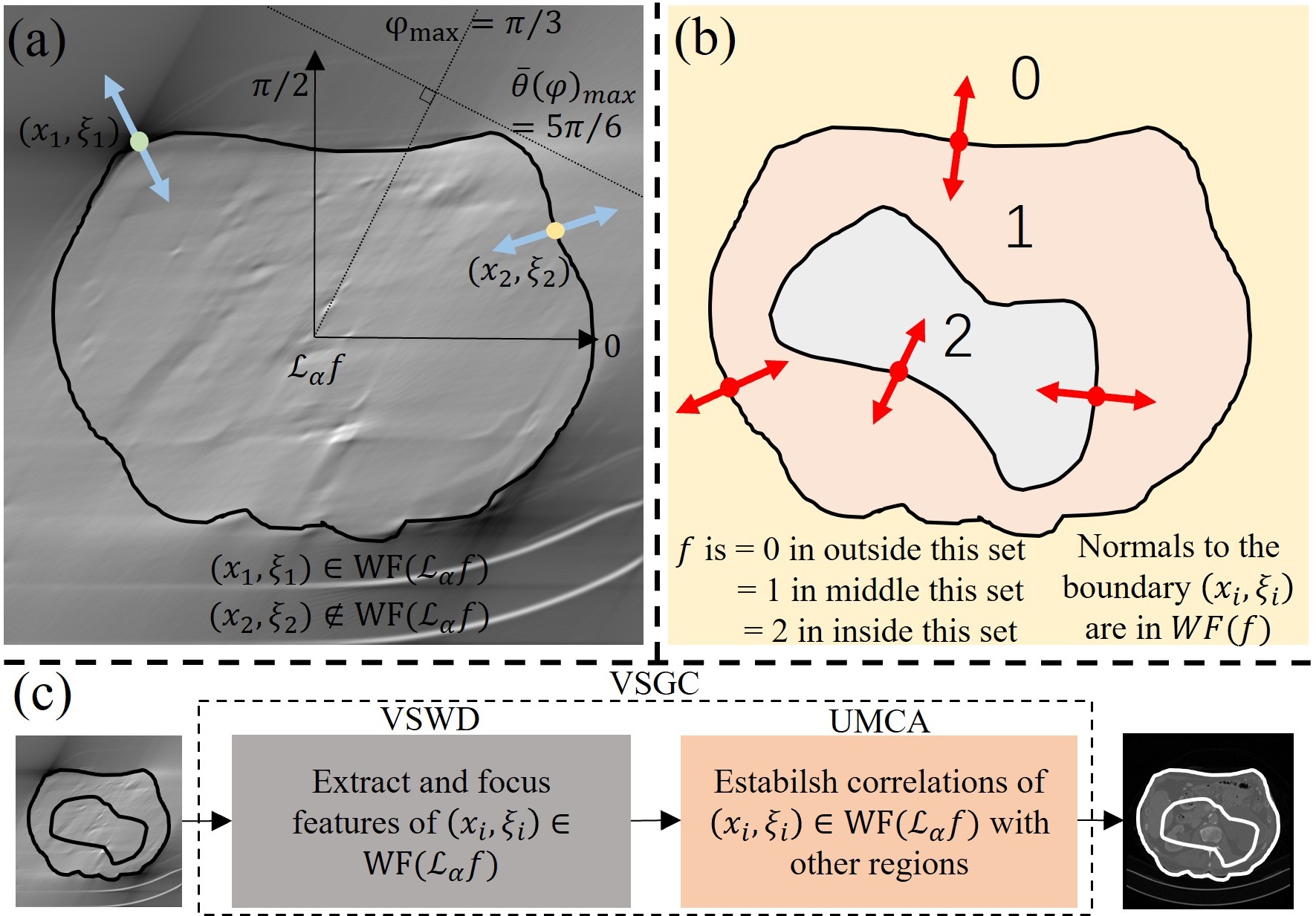} 
	\caption{Theoretical foundations and design rationale of VSGC: microlocal regularity, wavefront set interpretation. (a) Diagram of microlocal regularity theorem for $\mathcal{L}_\alpha$. (b) The boundary between 1 and 2 is assumed to be the boundary of the tissues within the body, and the boundary between 0 and 1 is assumed to be the boundary between the body and the air. This diagram explains what $WF(\cdot)$ is. (c) The design rationale of our model.} 
	\label{Motivation}
	\vspace{-15pt}
\end{figure}

As shown in Fig. \ref{Motivation}, $(x_1,\xi_1) \in \mathrm{WF}(\mathcal{L}_\alpha f)$ implies that $\mathcal{L}_\alpha f$ has a discontinuity at point $x_1$ along the direction $\xi_1$, corresponding to distinct edge features. $(x_2,\xi_2) \notin \mathrm{WF}(\mathcal{L}_\alpha f)$ indicates that $\mathcal{L}_\alpha f$ has no discontinuity at point $x_2$ along the direction $\xi_2$. We first designed the VSWD to focus on the structural information around $x_1$, for the VS edge features. Then the UMCA was designed to establish connections between these edge features and other regions of $\mathcal{L}_\alpha f$, by leveraging the links between the VS regions (where singularities are most completely preserved and structural information is most abundant) and other regions, so that the singularities and structural information of other regions (especially at the IVS) can be recovered. The combination of the UMCA and the VSWD resulted our proposed VSGC architecture.

\begin{figure*}[!t]
	\centering
	\includegraphics[width=0.9\textwidth]{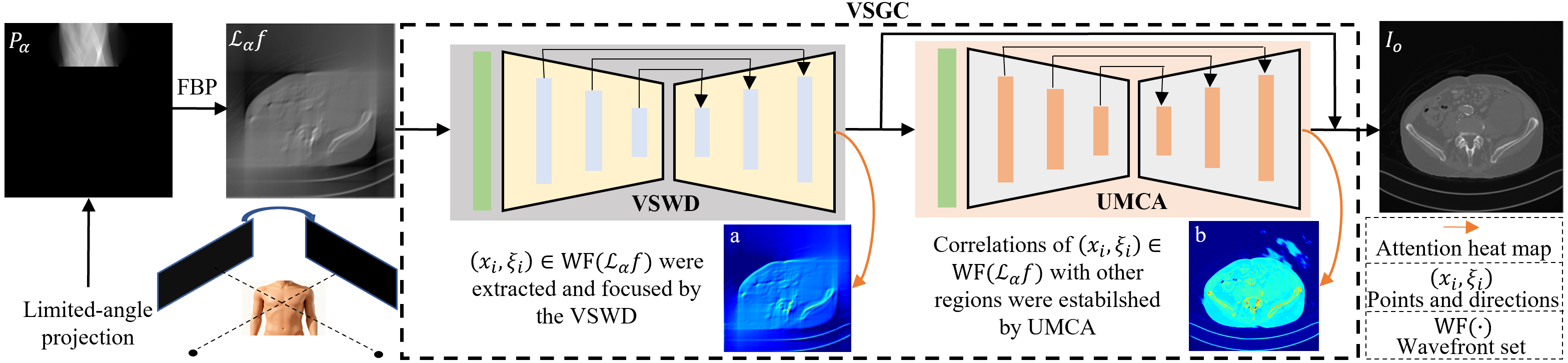}
	\caption{Overall design framework of VSGC. VSGC consists of VSWD and UMCA. The design goal of VSWD is to complete the extraction and focusing of VS edge features. Figure a shows the model attention heatmap output by VSWD, which focuses on VS edge features. The design goal of UMCA is to establish the connection between VS edge features and other regions of $\mathcal{L}_\alpha f$, and fully utilize VS edge features to recover the structural information of other regions. Figure b shows the model attention heatmap output by UMCA, which recovers the structural information of other regions while preserving the integrity of VS.}
	\label{VSGC}
	\vspace{-15pt}
\end{figure*}

\subsection{Network Architecture of VSGC}

As show in Fig. \ref{VSGC}, the proposed VSGC architecture is composed of VSWD and UMCA connected in series, with the former for the extraction and refinement of the VS edge features fed into the later. The UMCA adopts the MCA module that integrates the self-attention mechanism and convolution operation, to establish the correlations between VS edge features and other regions. In summary, the network can achieve three main objectives: the preservation of VS, the recovery of IVS, and the elimination of artifacts.

\subsubsection{Visible Singularities Wavelet Dense}

The design goal of the VSWD is to enhance the edge feature by representation modeling of VS. The core characteristics of VS stems from high-frequency edge structures, and its distinguishability can then be preliminarily enhanced by fusing the basic information of $\mathcal{L}_{a} f$ with directional edge features:
\begin{equation}
	F_{e} = \mathcal{L}_{a} f + \mathcal{L}_{a} f * G_{x} + \mathcal{L}_{a} f * G_{y},
\end{equation}
where $G_{x}$ and $G_{y}$ are operators for extracting edge features in the $x$-directions and $y$-directions (e.g., the Sobel operator), and $F_{e}$ denotes the edge-enhanced feature map.

\begin{figure}[!t]
	\centering
	\includegraphics[width=0.9\columnwidth]{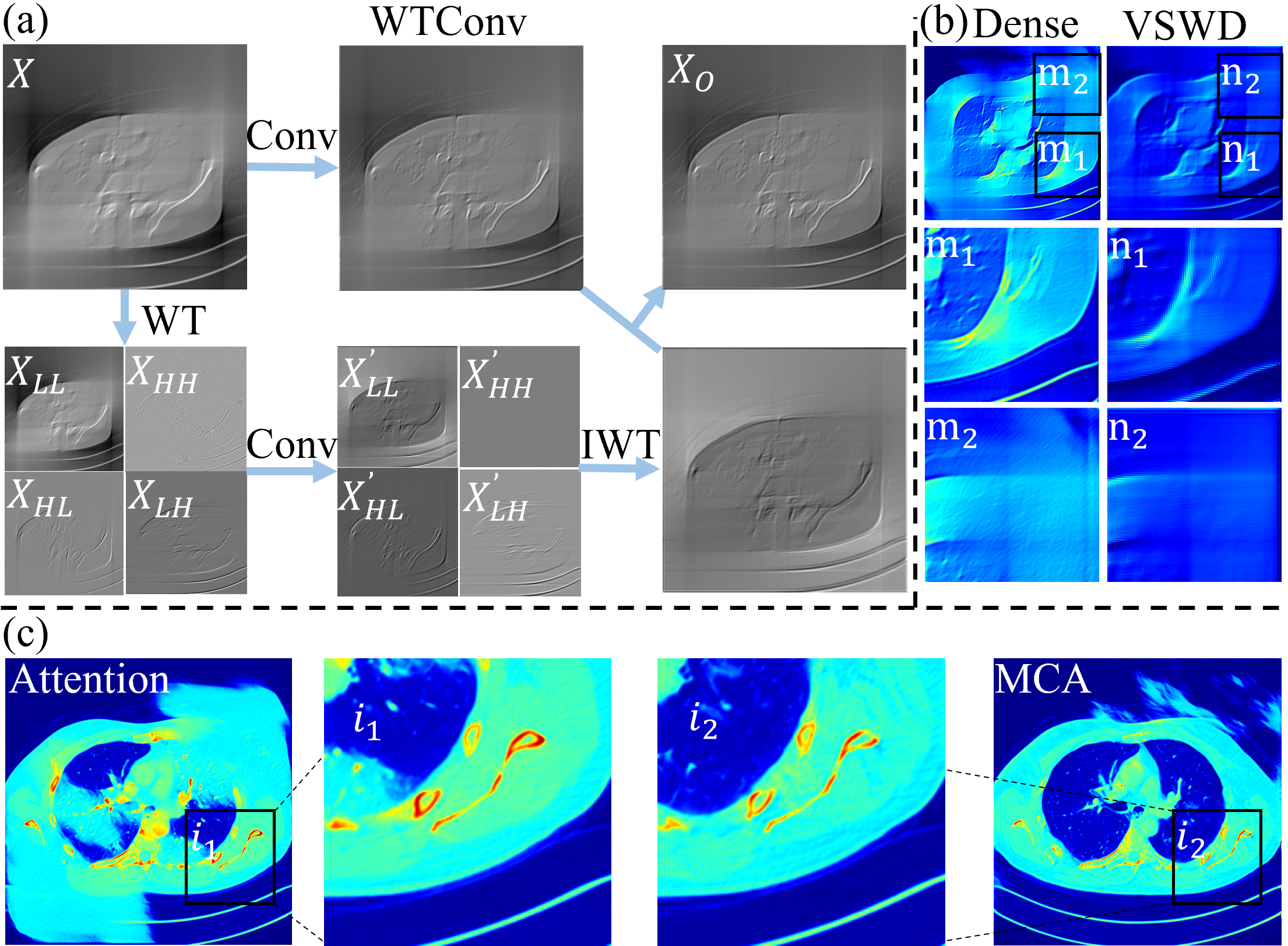} 
	\caption{WTConv workflow and attention heatmap comparisons. (a) The execution workflow of WTConv. (b) The attention heatmaps of the Dense \cite{huang2017densely} and the VSWD. VSWD focuses its attention on VS. (c) The attention heatmaps when a baseline attention\cite{vaswani2017attention} or UMCA is connected after the same VSWD pre-network. The structural detail level of $i_2$ is higher than that of $i_1$, and MCA achieves better restoration of the IVS, demonstrating its establishment of sufficient and reasonable correlations.}
	\label{WTconv_SAEL}
	\vspace{-15pt}
\end{figure}

Wavelet Transform can be applied to decompose $F_{e}$ into multi-scale subbands, which improves the separability between the high-frequency features of VS and the low-frequency features of IVS in the subbands. WTConv\cite{finder2024wavelet} was used to perform adaptive convolution on each subband in the wavelet domain to achieve accurate capture of the high-frequency features of VS. The wavelet subband decomposition can be described by:
\begin{equation}
	\left\{X_{LL}, X_{LH}, X_{HL}, X_{HH}\right\} = WT\left(F_{e}\right),
\end{equation}
where $WT(\cdot)$ denotes the 2D wavelet transform. $X_{LL}$ is the low-frequency subband (dominated by IVS), while $X_{LH}$, $X_{HL}$, and $X_{HH}$ are high-frequency subbands (dominated by VS). The subband convolution and reconstruction process is expressed as:
\begin{equation}
	\begin{aligned}
		&F_o = Conv(F_e,\mathcal{K}_0) + IWT({Conv}(X_{LL},\mathcal{K}_1) + \\
		&\quad Conv(X_{LH},\mathcal{K}_2) + Conv(X_{HL},\mathcal{K}_3) + Conv(X_{HH},\mathcal{K}_4)),
	\end{aligned}	
\end{equation}
where $IWT(\cdot)$ represents the inverse wavelet transform, and $\mathcal{K}_{i}$ ($i \in \{0,1,2,3,4\}$) are learnable convolution kernels. As illustrated in part (b) of Fig. \ref{WTconv_SAEL}, $Conv\left(X_{LH}, \mathcal{K}_{2}\right)$, $Conv\left(X_{HL}, \mathcal{K}_{3}\right)$, and $Conv\left(X_{HH}, \mathcal{K}_{4}\right)$ adaptively focus on the high-frequency subbands dominated by VS to enhance their structural features. The subband features are then reconstructed into the spatial-domain feature $F_{o}$ via the inverse wavelet transform, yielding accurate enhancement of the high-frequency features of VS.

\begin{figure*}[!t]
	\centering
	\includegraphics[width=0.9\textwidth]{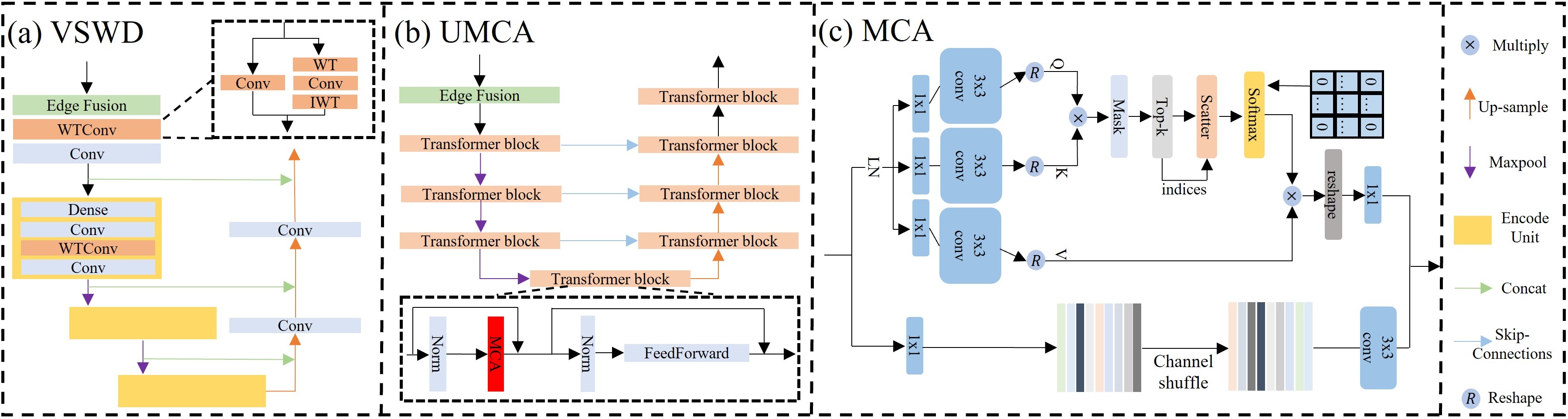} 
	\caption{VSGC core module architectures. (a) The architecture of VSWD. (b) The architecture of UMCA. (c) The architecture of MCA.}
	\label{model}
	\vspace{-15pt}
\end{figure*}

Multi-scale hierarchical encoding was used for integrating the structural information of VS across different scales, yielding more comprehensive representation of VS. The encoding and decoding process can be formulated as:
\begin{equation}
	x_1 = Conv(WTConv(F_e)),
\end{equation}
\begin{equation}
	x_i = E(x_{i-1}),\ i \in \{2,3,4\},
\end{equation}
\begin{equation}
	d_i = Conv(Cat(up(d_{i+1}), x_i)),\ i \in \{1,2,3\},
\end{equation}
where \( d_4 = x_4 \), \( up(\cdot) \) denotes upsampling, \( Cat(\cdot) \) denotes the concatenation of feature maps, and \( E(\cdot) \) is the encoding unit illustrated in Fig. \ref{model}. The feature tensor \( d_1 \) comprehensively preserves the high-frequency structural information of VS, facilitating the subsequent UMCA model to establish association modeling between VS and other regions.

\subsubsection{Multi-scale Visible Singularity-CrossRegion Correlation Self-Attention}

The fully connected baseline attention model can be described as follows:
\begin{equation}
	Attention(Q, K, V) = Softmax\left(\frac{QK^T}{\lambda}\right)V,
\end{equation}
where $Q$, $K$, and $V$ represent queries, keys, and values, respectively. $\lambda$ is the temperature parameter\cite{vaswani2017attention}. The MCA can be described by
\begin{equation}
	\begin{aligned}
		Attn_k &= \sum_{k=1}^{4} \alpha_k \cdot Softmax\left( M_k \odot \frac{QK^T}{\lambda} \right) V \\
		&+ Conv_{3} \left( CS \left( Conv_{2}(X) \right) \right),
	\end{aligned}
\end{equation}
where $M_k$ is a sparse mask that retains only the top k largest elements\cite{chen2023learning}. $\odot$ is elemental accumulation. $N$ is the total number of elements, $\quad k \in \{\frac{N}{2}, \frac{2N}{3}, \frac{3N}{4}, \frac{4N}{5}\}$. $\alpha_k$ is the learnable fusion weight satisfying $\sum_{k=1}^{4} \alpha_k = 1$. $X$ is the input feature map and $X \in \mathbb{R}^{1 \times C \times H \times W}$, $CS(\cdot)$ is the channel shuffling\cite{hu2024hybrid}, $Conv_{3}(\cdot)$ and $Conv_{2}(\cdot)$ are 3D convolution and 2D convolution respectively.

The role of the attention mechanism is to capture global or long-range dependencies among elements \cite{wang2018non}. Convolutions perform sliding computation on inputs via fixed-size kernels, adapted at local modeling \cite{he2016deep}. $Attn_k$ only retains the top-$k$ elements with the highest relevance, avoiding redundant computations for the baseline attention. It automatically focuses on the global dependencies of key structures (e.g., bones and organs) in the image as the key global correlations of edge features of VS in $\mathcal{L}_\alpha f$ (extracted and focused by VSWD). Channel shuffling can be used to establish detailed local correlations within image patches. The correlations established by MCA encompass the correlations between VS and other regions across most distances, eventually leading to the recovery of the singular values of these regions (particularly those at IVS).   

\subsubsection{Multi-Scale Loss Function with Anisotropic Constraints}
The perceptual loss $L_p$ is defined as:
\begin{equation}
	L_p(x,y) = \sum_{i=1}^5 \frac{1}{w_i h_i} \left\| RC_i(f(x)) - RC_i(y) \right\|^2,
\end{equation}
where $w_i, h_i$ denote the width and height of the input at layer $i$, respectively. $RC_i(\cdot)$ represents the $i$-th convolutional layer of the pre-trained RED-CNN\cite{chen2017low}. $f(x)$ is the predicted result of the VSGC and $y$ denotes the ground truth. The anisotropic weighted loss $L_a$ is defined as:
\begin{equation}
	L_a(x,y) =  \left\| W \cdot (f(x) - y) \right\|^2,
\end{equation}
where $W$ is a matrix of the same size as the output, which takes a value of 2 at IVS, and 1 at VS. The regions corresponding to IVS rotate around the matrix center as the limited-angle range changes, enabling parameter-update with larger weights for regions severely affected by artifacts. The SSIM loss $L_s$ is defined as:
\begin{equation}
	L_s(x,y) = 1 - \frac{(2\mu_{\hat{y}}\mu_y + c_1)(2\sigma_{\hat{y}y} + c_1)}{(\mu_{\hat{y}}^2 + \mu_y^2 + c_2)(\sigma_{\hat{y}}^2 + \sigma_y^2 + c_2)},
\end{equation}
where $\mu_{\hat{y}}$ denotes the pixelwise mean of the model prediction, $\mu_y$ denotes the pixel mean of the ground truth, $\sigma_{\hat{y}y}$ denotes the covariance of the model predicted and labeled values, and $\sigma_{\hat{y}}^2$ and $\sigma_y^2$ denote the pixelwise variance of the model prediction and the ground truth, accordingly. The edge gradient loss $L_e$ is defined as:
\begin{equation}
	L_e(x,y) = \left\| Grad(f(x)) - Grad(y) \right\|^2,
\end{equation}
where $Grad(\cdot)$ denotes the gradient operator. All loss terms are weighted and summed, with parameters $\lambda, \alpha, \beta, \gamma$ controlling the relative weights of each loss function:
\begin{equation}
	L(x,y) = \lambda L_p(x,y) + \alpha L_s(x,y) + \beta L_e(x,y) + \gamma L_a(x,y).
\end{equation}
The proposed multi-scale loss function can enhance image texture detail, edge sharpness, and human visual perception, effectively addressing image smoothing issues\cite{han2022perceptual}\cite{ma2020low}\cite{gholizadeh2020deep}.


\section{Experiments}
\label{experiments}

\subsection{Data Specification}

\subsubsection{AAPM Challenge Data}

The dataset utilized for model training and testing in this study were provided by the Mayo Clinic. 4,742 slices derived from 9 patients were allocated for training, while 100 randomly selected slices from an additional patient were designated for testing. Each slice has a thickness of 1 mm. Projection data were generated by simulation using a ray-driven algorithm \cite{lee2019explainable, bi2019artificial}. The Source-to-Detector Distance (SDD) from the X-ray source to the detector panel is 60 cm, and the Source-to-Orbit Distance (SOD) from the X-ray source to the rotation center is 40 cm. The detector array consists of 720 units with a total width of 41.3 cm.


\subsubsection{Periapical Data}

\begin{figure}[!t]
	\centering
	\includegraphics[width=0.9\columnwidth]{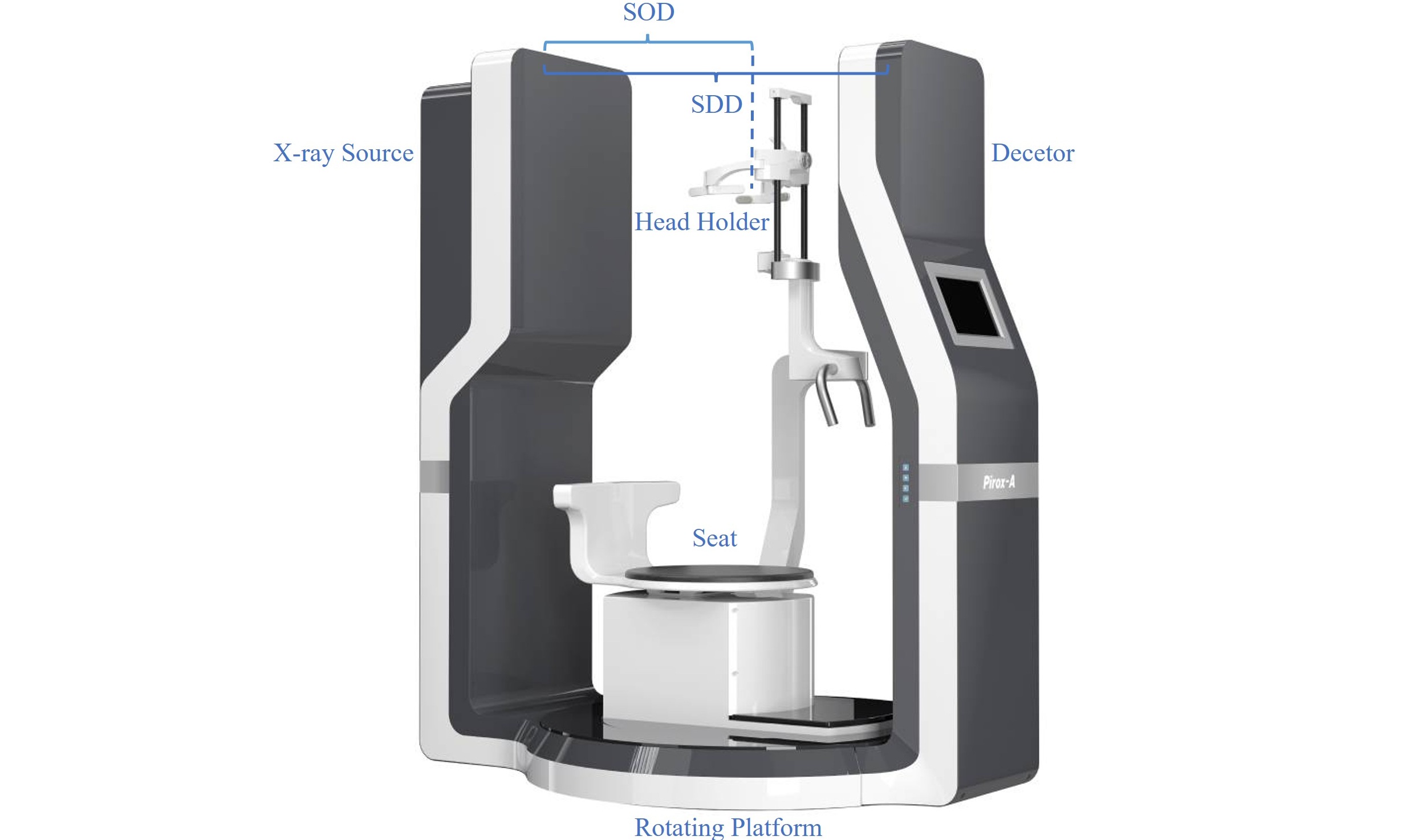} 
	\caption{The Pirox CT system, designed and manufactured by YOFO Medical Technology Co., Ltd., with its scanning geometry indicated.}
	\label{Priox}
	\vspace{-15pt}
\end{figure}

The periapical dataset is clinical data acquired by a Pirox dental cone beam computed tomography(CBCT) device designed and manufactured by YOFO (Hefei) Medical Technology Co., Ltd. As show in Fig. \ref{Priox}, such device has a SOD of 350 mm and a SDD of 600 mm. The image resolution is 0.2 mm per pixel. It operates at a voltage of 100 kV and a current of 8 mA. There are 20 cases in the dataset, each containing 360 slices. The dataset initially comprises full-view projections and their corresponding FDK-reconstructed volumetric images, with a size of 640 × 640 × 360. We extract limited-angle projections from the full-view data, perform FDK reconstruction to generate volumetric data, and then select all 128 × 128 slices located at the positions of the left and right dental arches from the reconstructed volume to form the paired dataset.

\subsubsection{Mouse Data}

This study adopted an actual raw projection dataset of annotated mice acquired in June 2025 by the Jinan Laboratory of Applied Nuclear Science. The dataset was collected via a MARS multi-energy CT system, covering two energy ranges (21–30 kVp and 31–60 kVp), and comprised 2,835 slices. All data were preprocessed to ensure consistency with the training dataset. For experimental evaluation, one slice was randomly selected from each anatomical region, yielding a total of 50 slices. The scanning parameters were as follows: the SOD was 76.28 mm, the SDD was 361.1 mm, and the detector pixel size was 0.1 mm.

\subsection{Implementation Details}

In this study, all experiments were conducted on a high-performance workstation equipped with two NVIDIA RTX 4090D. The Adam optimizer was used with a learning rate of $lr=2.5\times10^{-4}$. The hyperparameters of the loss function were configured as $\alpha = 1.0$, $\beta = 0.1$, $\lambda = 0.5$, $\mu = 0.3$ and $\gamma = 0.3$. To comprehensively verify the validity and feasibility of the proposed model design rationale, we conducted comparative experiments with seven representative methods: FBP\cite{kak2001principles}, RED-CNN\cite{chen2017low}, FBPConvnet\cite{jin2017deep}, WISM\cite{zhang2024wavelet}, MIST-Net\cite{pan2022multi}, DIOR\cite{hu2022dior} and IRON\cite{pan2024iterative}. In particular, the three iterative methods MIST-Net, DIOR and IRON were all run with 30 iterations, and the diffusion-based method WISM was implemented with a sampling step size of 200. Two widely used quantitative metrics were employed for performance assessment: peak signal-to-noise ratio (PSNR) and structural similarity (SSIM).


\subsection{Reconstruction Experiments}

\subsubsection{AAPM Data Results}

\begin{figure*}[!t]
	\centering
	\includegraphics[width=0.9\textwidth]{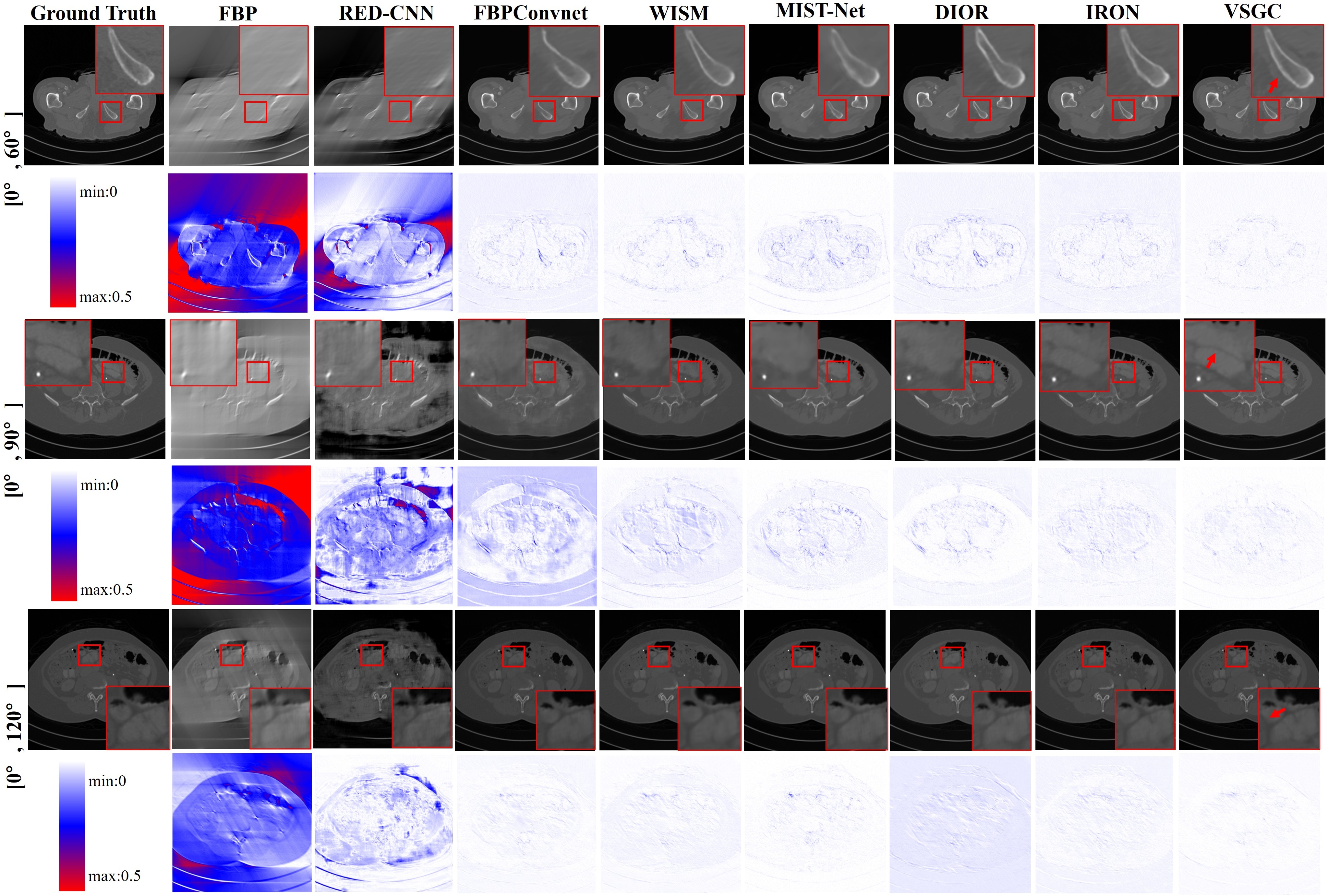} 
	\caption{Comparison of reconstruction results between VSGC and other methods under different angular ranges on the AAPM simulation dataset. Rows 1 and 2 correspond to the angular range of [0°, 60°]; rows 3 and 4 correspond to [0°, 90°]; rows 5 and 6 correspond to [0°, 120°]. Among them, rows 2, 4 and 6 are the differences between the reconstruction results of the corresponding methods and the ground truth.}
	\label{aapm}
	\vspace{-5pt}
\end{figure*}     

\begin{figure*}[!t]
	\centering
	\includegraphics[width=0.9\textwidth]{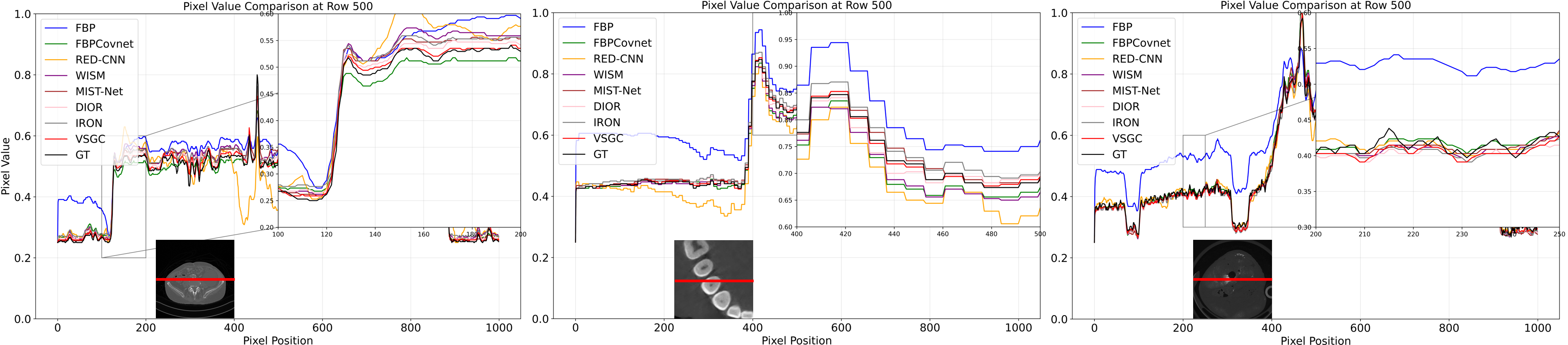} 
	\caption{Reconstructed intensity distribution map of the specified red line in the image. From left to right: aapm simulation dataset , clinical periapical CBCT dataset, mouse dataset.}
	\label{line}
	\vspace{-15pt}
\end{figure*}

As shown in Fig. \ref{aapm}, we randomly selected one representative case for each angular range from the AAPM simulated LACT test set for the visual demonstration. For the [0°, 60°] limited-angle scenario, the bone morphology failed to be faithfully recovered by FBP and RED-CNN. Among the remaining competing methods, the bone morphology from the VSGC is the closest to the ground truth with well-restored edge details. For the [0°, 90°] scenario, all methods except IRON and VSGC failed to reconstruct the tissue boundaries accurately. In comparison to IRON, VSGC recovered clearer tissue boundaries. For the [0°, 120°] scenario, the intestinal folds recovered by VSGC were the most consistent with the ground truth, while some fine structures failed be restored by the other competing methods. Table \ref{aapm_table} presents the quantitative comparison of reconstruction results obtained by different methods on the AAPM simulated dataset for various angular ranges. VSGC achieved considerable improvements over IRON at both [0°, 90°] and [0°, 120°] scenarios. Notably, at the [0°, 60°] case, VSGC still maintained high reconstruction metric values, whereas they dropped drastically for the other methods. As illustrated in Fig. \ref{line}, analysis of the intensity profile along the red line within the AAPM simulated dataset indicates that the intensity curve generated by VSGC exhibits almost perfect coincidence with the ground truth. In contrast, the results from the other methods fluctuate around the ground truth with obvious deviations. The abovementioned results confirmed the outstanding edge-preserving capability of the proposed VSGC.

\begin{table}[ht]
	\centering
	\small
	\setlength{\tabcolsep}{3pt}
	\caption{Comparison of Quantitative Reconstruction Results\\ Among Different Methods on the AAPM Simulation Dataset}
	\label{aapm_table}
	\begin{tabular}{c|cc|cc|cc}
		\toprule
		\multirow{2}{*}{Methods} & \multicolumn{2}{c|}{$[0^\circ, 60^\circ]$} & \multicolumn{2}{c|}{$[0^\circ, 90^\circ]$} & \multicolumn{2}{c}{$[0^\circ, 120^\circ]$} \\
		\cmidrule(lr){2-3} \cmidrule(lr){4-5} \cmidrule(lr){6-7}
		& PSNR & SSIM & PSNR & SSIM & PSNR & SSIM \\
		\midrule
		FBP\cite{ramachandran1971three} & 10.65 & 0.4408 & 11.76 & 0.3386 & 14.80 & 0.4332 \\
		RED-CNN\cite{chen2017low} & 15.73 & 0.6009 & 19.50 & 0.6506 & 24.87 & 0.7663 \\
		FBPConvnet\cite{jin2017deep} & 32.46 & 0.9315  & 36.70 & 0.9530 & 38.80 & 0.9685 \\ 
		MIST-Net\cite{pan2022multi} & 34.25 & 0.9425 & 37.98 & 0.9528 & 38.99 & 0.9691\\
		WISM\cite{zhang2024wavelet} & 36.39 & 0.9528 & 38.72 & 0.9686 & 39.41 & 0.9747 \\
		DIOR\cite{hu2022dior} & 35.86 & 0.9483 & 38.95 & 0.9645 & 39.11 & 0.9717\\
		IRON\cite{pan2024iterative}  & 36.92 & 0.9571 & 39.59 & 0.9652 & 40.78 & 0.9755 \\
		\textbf{VSGC}  & \textbf{39.37} & \textbf{0.9720}   & \textbf{40.44} & \textbf{0.9764} & \textbf{41.26} & \textbf{0.9808} \\
		\bottomrule
	\end{tabular}
\end{table}

\subsubsection{Periapical Data Results}

\begin{figure*}[!t]
	\centering
	\includegraphics[width=0.9\textwidth]{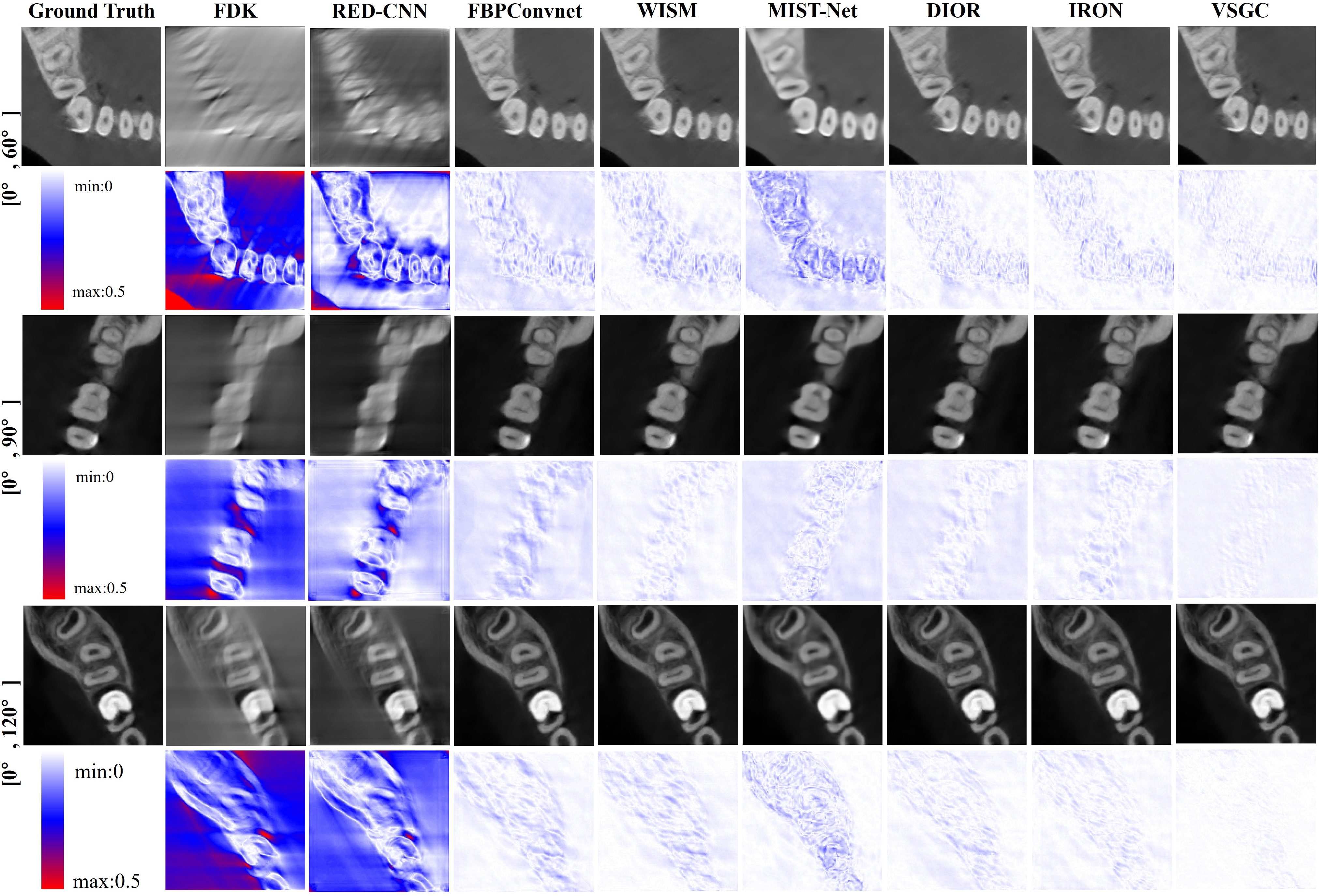} 
	\caption{Comparison of reconstruction results between VSGC and other methods under different angular ranges on the clinical periapical CBCT dataset. Rows 1 and 2 correspond to the angular range of [0°, 60°]; rows 3 and 4 correspond to [0°, 90°]; rows 5 and 6 correspond to [0°, 120°]. Among them, rows 2, 4 and 6 are the differences between the reconstruction results of the corresponding methods and the ground truth.}
	\label{yapian}
	\vspace{-5pt}
\end{figure*}

As shown in Fig. \ref{yapian}, we randomly selected one representative case for each angular range from the LACT test set of the clinical periapical CBCT dataset for visualization. At [0°, 60°], the reconstructions generated by FDK and RED-CNN exhibit severe streak artifacts. In contrast, results of other competing methods show similarities in dental anatomical structures, yet exhibit discrepancies in CT number quantification. By comparing the [0°, 60°] reconstructions of competing methods against the ground truth, it is evident that VSGC exhibits the smallest discrepancies. For the [0°, 90°] and [0°, 120°] scenarios, the difference maps of VSGC can be observed as nearly uniformly white, containing only scattered light-blue regions and no distinct structural mismatches, which confirms the superior capability of VSGC in reconstructing fine dental anatomical structures. Table \ref{yapian_table} presents quantitative assessments of reconstruction performance across competing methods on the clinical periapical CBCT dataset under various angular ranges. The proposed VSGC attains notably high quantitative reconstruction metrics at [0°, 90°] and [0°, 120°]. With deteriorated performance at [0°, 60°], VSGC still outperforms all competing methods. Fig. \ref{line} depicts the intensity distribution along the red line in the clinical periapical CBCT dataset: the red curve represents the outputs of VSGC, whereas the black curve denotes the ground truth, and their high consistency confirms the enhanced reconstruction performance of the VSGC.

\begin{table}[ht]
	\centering
	\small  
	\setlength{\tabcolsep}{3pt}  
	\caption{Comparison of Quantitative Reconstruction Results Among Different Methods Using Clinical Periapical CBCT Datasets}
	\label{yapian_table}
	\begin{tabular}{c|cc|cc|cc}
		\toprule
		\multirow{2}{*}{Methods} & \multicolumn{2}{c|}{$[0^\circ, 60^\circ]$} & \multicolumn{2}{c|}{$[0^\circ, 90^\circ]$} & \multicolumn{2}{c}{$[0^\circ, 120^\circ]$} \\
		\cmidrule(lr){2-3} \cmidrule(lr){4-5} \cmidrule(lr){6-7}
		& PSNR & SSIM & PSNR & SSIM & PSNR & SSIM\\
		\midrule
		FBP\cite{ramachandran1971three} & 13.23 & 0.4030 & 14.20 & 0.4709 & 17.04 & 0.6169 \\
		RED-CNN\cite{chen2017low} & 18.23 & 0.7126 & 20.12 & 0.7912 & 24.53 & 0.8243 \\
		FBPConvnet\cite{jin2017deep} & 32.66 & 0.9582  & 33.13 & 0.9662 & 34.45 & 0.9748 \\
		MIST-Net\cite{pan2022multi} & 29.87 & 0.9458 & 38.54 & 0.9640 & 39.49 & 0.9644\\
		WISM\cite{zhang2024wavelet}  & 32.67 & 0.9590 & 40.13 & 0.9795 & 41.17 & 0.9815 \\
		DIOR\cite{hu2022dior} & 33.72 & 0.9662 & 40.46 & 0.9818 & 41.57 & 0.9833\\
		IRON\cite{pan2024iterative}  & 33.30 & 0.9647 & 41.29 & 0.9896 & 42.44 & 0.9903 \\
		\textbf{VSGC} & \textbf{37.34} & \textbf{0.9855} & \textbf{42.83} & \textbf{0.9944} & \textbf{44.26} & \textbf{0.9954} \\
		\bottomrule
	\end{tabular}
\end{table}


\subsubsection{Mouse Data Results}

\begin{figure*}[!t]
	\centering
	\includegraphics[width=0.9\textwidth]{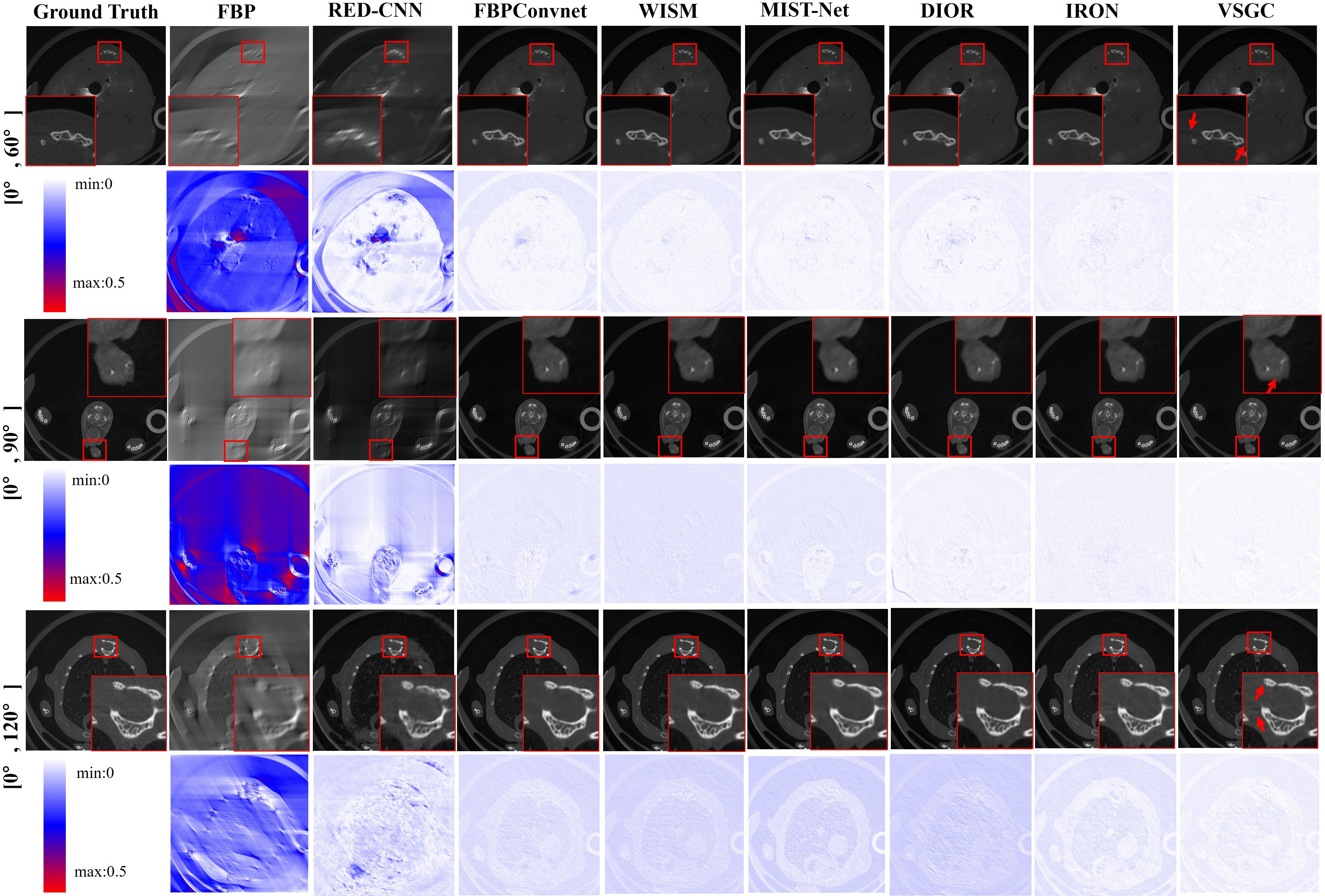} 
	\caption{Comparison of reconstruction results between VSGC and other methods under different angular ranges on the real mouse dataset. Rows 1 and 2 correspond to the angular range of [0°, 60°]; rows 3 and 4 correspond to [0°, 90°]; rows 5 and 6 correspond to [0°, 120°]. Among them, rows 2, 4 and 6 are the differences between the reconstruction results of the corresponding methods and the ground truth.}
	\label{mouse}
	\vspace{-15pt}
\end{figure*}

As shown in Fig. \ref{mouse}, we randomly selected one representative case per angular range from the LACT test subset of the mouse dataset for qualitative visualization. For the [0°, 60°] scenario, FBP and RED-CNN failed to faithfully recover the bone morphological features. The proposed VSGC exhibits fine black fissures on both sides of the bone, a subtle structural feature failed to be recovered by competing methods. For the [0°, 90°] scenario, the VSGC was the only method successfully reconstructed the calcification foci at the regions indicated by red arrows. For the [0°, 120°] scenario, VSGC’s reconstruction clearly revealed three distinct intraosseous voids at the regions denoted by red arrows, whereas the leftmost void from other methods appears notably blurred and poorly delineated. Table \ref{xiaoshu_120_table} presents quantitative comparisons for the mouse dataset. Due to severe ring artifacts and high noise levels in this dataset, most models only learnt general structural features, and failed to capture irregular noise components. Consequently, no statistically significant performance disparity is observed between VSGC and other competing methods. In addition, the intensity profiles along the red lines in the mouse dataset indicate that all methods exhibit deviations from the ground truth due to noise interference, as shown in Fig. \ref{line}.

\begin{table}[ht]
	\centering
	\small  
	\setlength{\tabcolsep}{3pt}  
	\caption{Comparison of Quantitative Reconstruction Results Across Different Methods Using Mouse Datasets} 
	\label{xiaoshu_120_table}
	\begin{tabular}{c|cc|cc|cc}
		\toprule
		\multirow{2}{*}{Methods} & \multicolumn{2}{c|}{$[0^\circ, 60^\circ]$} & \multicolumn{2}{c|}{$[0^\circ, 90^\circ]$} & \multicolumn{2}{c}{$[0^\circ, 120^\circ]$} \\
		\cmidrule(lr){2-3} \cmidrule(lr){4-5} \cmidrule(lr){6-7}
		& PSNR & SSIM & PSNR & SSIM & PSNR & SSIM\\
		\midrule
		FBP\cite{ramachandran1971three} & 12.88 & 0.4882 & 11.09 & 0.4553 & 15.21 & 0.5757 \\ 
		RED-CNN\cite{chen2017low} & 22.39 & 0.6935 & 22.56 & 0.7242 & 24.89 & 0.7758 \\ 
		FBPConvnet\cite{jin2017deep} & 35.03 & 0.9152 & 35.01 & 0.9183 & 35.63 & 0.9227\\ 
		MIST-Net\cite{pan2022multi} & 36.10 & 0.9196  & 36.61 & 0.9241 & 37.07 & 0.9291 \\ 
		WISM\cite{zhang2024wavelet}  & 36.77 & 0.9247 & 37.20 & 0.9282 & 37.35 & 0.9314 \\ 
		DIOR\cite{hu2022dior} & 35.60 & 0.9156 & 36.64 & 0.9241 & 37.18 & 0.9289\\
		IRON\cite{pan2024iterative}  & 36.41 & 0.9203 & 36.87 & 0.9256 & 37.59 & 0.9375 \\ 
		\textbf{VSGC} & \textbf{37.09} & \textbf{0.9313} & \textbf{37.49} & \textbf{0.9344} & \textbf{37.70} & \textbf{0.9379} \\ 
		\bottomrule
	\end{tabular}
\end{table}

\subsection{Computational Complexity of VSGC}

Table \ref{Costing and testing time} presents a comparison of the model parameters and computational complexity across different methods. The number of training iterations for MIST-Net, DIOR, and IRON was uniformly set to 30, while the number of diffusion sampling steps for WISM was set to 200. A total of 100 2D slices were randomly selected as the test set. Notably, the inference time of VSGC is reduced to approximately 1/3 of IRON’s. VSGC has only 10.5 M model parameters.

\subsection{High-Resolution LACT}

\begin{figure}[!t]
	\centering
	\includegraphics[width=\columnwidth]{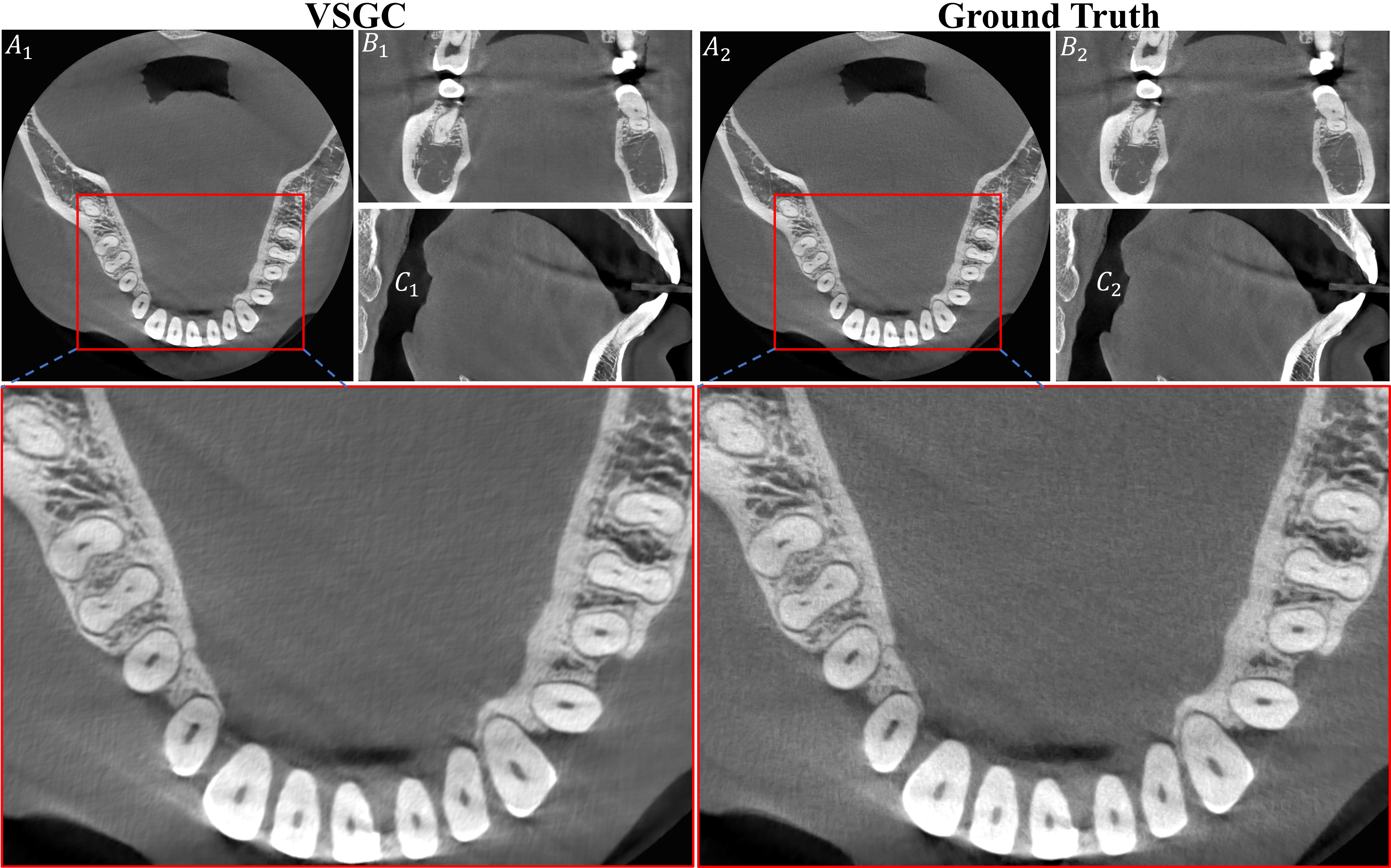} 
	\caption{High-resolution LACT reconstruction results. $A_i, B_i, C_i(i \in {1,2})$ represent the axial, coronal, and sagittal planes, respectively. The regions enclosed by red boxes are magnified and presented at the bottom.}
	\label{PA_High}
	\vspace{-15pt}
\end{figure}

The proposed VSGC was trained on a clinical dataset acquired using a device (illustrated in Fig. \ref{Priox}) manufactured by YOFO Medical and operated in its high-resolution mode. The acquired images had a spatial resolution of 0.1 mm per pixel and a size of 1000×1000 pixels. In [0°, 90°] angular range, VSGC achieved a mean PSNR of 39.16 dB and a mean SSIM of 0.9742 in quantitative terms. Notably, the reconstruction time for a single high-resolution slice was only 0.235 seconds. A randomly selected example of high-resolution LACT reconstruction results by VSGC is shown in Fig. \ref{PA_High}. Thanks to its lightweight design, VSGC enables high-resolution imaging reconstruction under the constraint of limited hardware resources (24 GB GPU memory).


\subsection{Ablation Study}

\begin{figure}[!t]
	\centering
	\includegraphics[width=0.9\columnwidth]{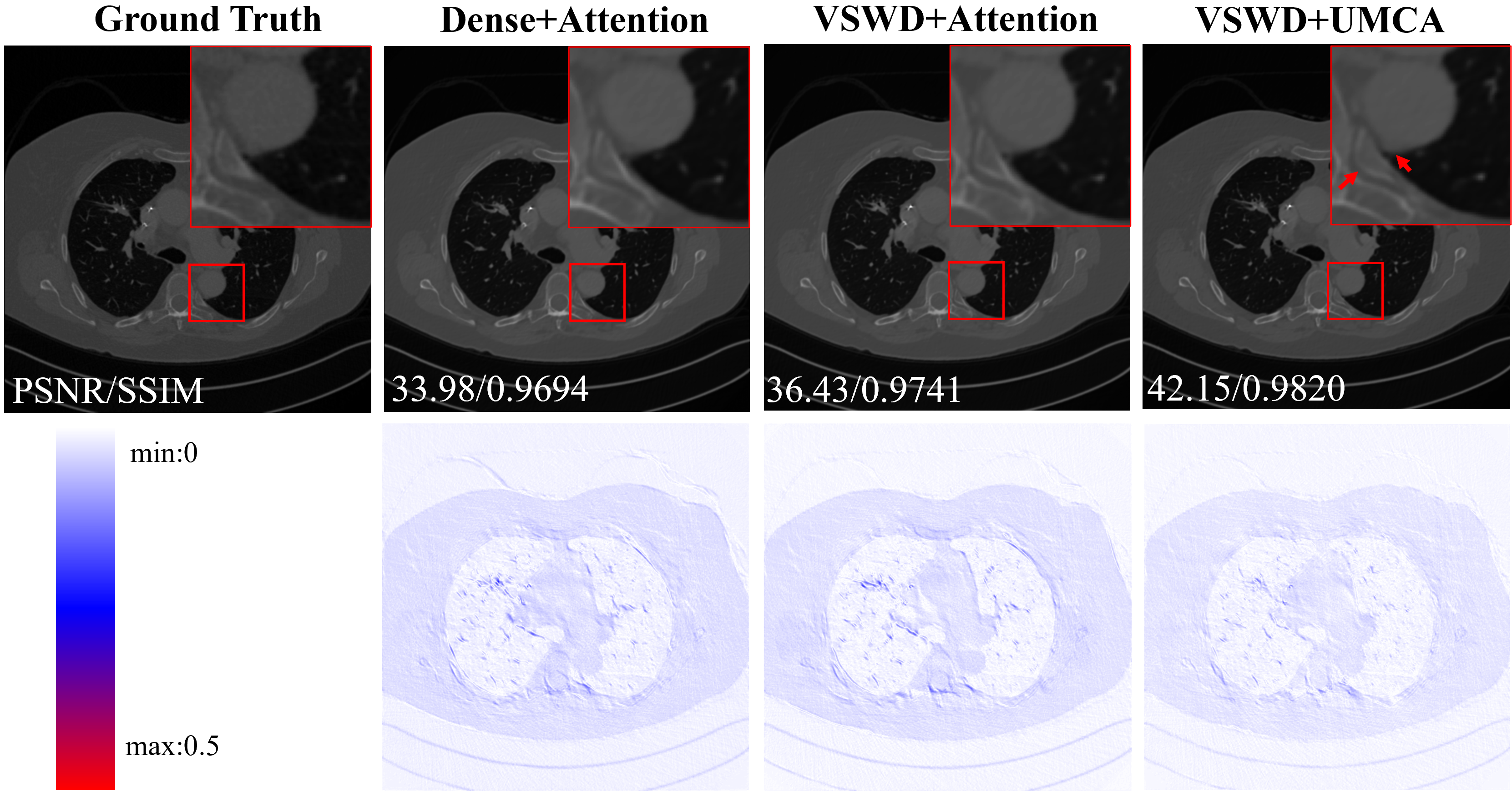} 
	\caption{Comparison of reconstruction results for different model combinations. Dense model reference \cite{huang2017densely}, Attention model reference baseline attention \cite{vaswani2017attention}.}
	\label{attention_map}
	\vspace{-15pt}     
\end{figure}

As shown in Fig. \ref{attention_map}, ablation experiments were performed by replacing VSWD and UMCA with a dense network\cite{huang2017densely} and the baseline attention mechanism\cite{vaswani2017attention}, with respectively, to validate the efficacy of the proposed design. Compared to the Dense+Attention combination, VSWD+Attention yields reconstruction results with significantly sharper bone-soft tissue boundaries. Replacing the baseline attention mechanism with UMCA enhances the comprehensiveness and structural consistency of feature correlations, enabling the model to restore finer anatomical details. Overall, the proposed VSGC conforms to the design rationale and exhibits superior performance in tissue edge reconstruction. This finding is further verified by analyzing edge structural information in residual maps, where the VSWD+UMCA combination contains the least edge feature residues, confirming the effectiveness of the proposed components in reducing structural discrepancies. Quantitative comparisons of reconstruction results among different model combinations are listed in Table \ref{ablation_table}, which further validates the effectiveness of the proposed design.

\begin{table}[h]
	\centering
	\caption{Quantitative Comparison of Calculated Costs}
	\vspace{-5pt}
	\begin{tabular}{cccccc}
		\toprule
		Methods  & WISM & MIST-Net & DIOR & IRON & \textbf{VSGC}  \\
		\midrule
		Cost time(s) & 2016.45 & 45.27 & 219.23 & 66.48 & \textbf{23.69} \\
		Parameters(M) & 36.1 & 20.4 & 12.4 & 27.1 & 
		\textbf{10.5}\\
		\bottomrule
	\end{tabular}
	\label{Costing and testing time}
	\vspace{-10pt}
\end{table}

\begin{table}[h]
	\centering
	\caption{Quantitative Comparison of Model Validity}
	\vspace{-5pt}
	\begin{tabular}{ccc}
		\toprule
		Methods & PSNR & SSIM \\
		\midrule
		Dense+Attention & 37.25 & 0.9637 \\
		VSWD+Attention & 39.87 & 0.9703\\
		VSWD+MCA & \textbf{40.44} & \textbf{0.9764}\\
		\bottomrule
	\end{tabular}
	\label{ablation_table}
	\vspace{-10pt}
\end{table}

\section{Discussion}
\label{discussion}

VSGC leverages the correlations between high-fidelity singularities in the VS region and those in other regions to recover singularities in IVS regions, which is fundamentally distinct from comparative methods relying on extra priors, expanded data domains, or data consistency exploration. Rooted in the underlying mathematical principles of LACT reconstruction, VSGC fully accounts for LACT reconstruction anisotropy and mines structural information of the VS region that closely matches the ground truth, thus boosting reconstruction performance. This provides a novel solution for developing networks targeting images with LACT-like characteristics.
VSGC presents three key advantages: it completes the prediction of a 360-slice LACT volume in only 1 minute; its reconstruction performance under the [0°, 60°] angular range outperforms comparative methods significantly; its lightweight design enables high-resolution LACT imaging. Despite outperforming other comparative methods, this approach has limitations. Specifically, a self-attention mechanism is adopted to establish correlations between VS edge features and those in other regions, and this algorithm suffers from high time complexity. Future work will focus on improving VSGC’s execution efficiency. However, it should be emphasized that authenticity is paramount in medical image reconstruction, as false images may compromise the accuracy of clinical diagnosis.


\section{Conclusion}  
\label{conclusion}

This work considered the core imaging characteristics of LACT arising from insufficient scanning angles, including the directionality of artifacts and the directional loss of structural information—intrinsic properties of LACT that align with the well-established VS and IVS theories. Based on these theories, we proposed a new paradigm for the LACT reconstruction: extract and focus on enhanced VS edge features, followed by the learning of targeted correlations between VS and other image regions. Under such paradigm, we proposed a novel network architecture (VSGC) composed of a dedicated model component (VSWD) for the edge feature extraction and a modified attention mechanism (MCA) targeting the correlations between VS and other regions. The results obtained from extensive experiments on both simulated and clinical datasets suggested the feasibility and effectiveness of the proposed design paradigm and architecture.


\appendices
\bibliographystyle{IEEEtran}
\justifying
\bibliography{sample,reference}

\end{document}